\documentclass[11pt]{article}
\input epsf.sty
\usepackage{epsfig} %,showkeys}

\begin{document}

{}~ \hfill\vbox{\hbox{hep-th/mmyyxxx}

\hbox{HUTP-05/A0032}

 }\break

\vskip 1cm

\begin{center}
\Large{\bf Re-Recounting Dyons in N=4 String Theory}

\vspace{0.6cm}

\vspace{20mm}

\large{Davide Gaiotto$^{a}$}

\vspace{10mm}

\large{\em $^{a}$ Jefferson Physical Laboratory, Harvard
University,}

\vspace{0.2cm} \normalsize{\em Cambridge, MA 01238, USA}

\vspace{0.4cm}

\end{center}

\vspace{10mm}

\begin{abstract}

\bigskip
The purpose of this brief note is to understand the reason for the
appearance of a genus two Riemann surface in the expression for the
microscopic degeneracy of $1/4$ BPS dyons in N=4 String Theory.
\medskip

\end{abstract}

\newpage
$N=4$ string theory has a duality group

\begin{equation}
SL(2,Z) \times O(22,6,Z)
\end{equation}

The $U(1)$ charges of the theory form a $SL(2,Z)$ doublet ${q_e
\choose q_m}$ of $O(22,6,Z)$ vectors. Objects with $q_e = 0$ or $q_m
= 0$ are half BPS, while $1/4$ BPS dyons may have generic charges.
The degeneracy of dyonic states depends only on the $O(22,6,Z)$
invariants

\begin{equation}
N_e = \frac{q_e^2}{2} \qquad N_{em} = q_e \cdot q_m \qquad N_m =
\frac{q_m^2}{2}
\end{equation}

The generating function for the degeneracies is conjectured to be
\cite{Dijkgraaf:1996it}

\begin{equation}
\sum_{N_e,\;N_m,\;N_{em}} e^{2 \pi i (N_e \rho + N_{em} v + N_m
\sigma)}d(N_e,N_{em},N_m) = \Phi\left[{\rho \, v \choose v \,
\sigma}\right]
\end{equation}

Here $\Phi[\Omega]$ is the partition function of $24$ chiral bosons
on a genus two Riemann surface of period matrix $\Omega$. This is
also the inverse of the unique automorphic form of weight $10$. The
conjecture has been verified in various ways (for example
\cite{Shih:2005uc}), here we want to understand more directly why a
genus two Riemann surface would appear in computing the partition
function for $1/4$ BPS states.

In the following we will use the realization of $N=4$ string theory
as IIB superstring theory compactified on $K3 \times T^2$. $26$ of
the possible $U(1)$ charges correspond to various branes and strings
wrapped along one of the $S^1$ and on cycles of the $K3$. Two more
electric charges correspond to momentum along the other circle and
one KK monopole charge. We will not turn on these two charges,
although it is possible to do so and repeat the following
construction with minor modifications.

If the $K3$ is much smaller than the $T^2$, a $1/2$ BPS state with
electric charges appears as a black string wrapped on one cycle of
the torus. This string is a bound state of $D5$, $NS5$ branes
wrapped on $K3$, $D3$ branes wrapped on the $22$ $2$-cycles of $K3$,
$D1$,$F1$ strings.

A black string wrapped on the other cycle of the torus corresponds
instead to a $1/2$ BPS magnetic object. The partition function for
such $1/2$ BPS objects is known to be

\begin{equation}
\sum d(N_e) e^{2 \pi i N_e \tau} = \frac{1}{\eta(\tau)^{24}}
\end{equation}

This is the torus partition function of $24$ chiral bosons. Indeed
in the low energy limit, the CFT living on the black string is the
same as the one for a fundamental heterotic string compactified on a
$T^6$ \cite{Hull:1994ys,Witten:1995ex}. The electric charges form a
Narain lattice of signature $(22,6)$. A supersymmetric ground state
for the string has level $0$ for the right movers, level matching
requires a level $N_e$ for the left mover oscillators. The partition
function is then the partition function for the $24$ left moving
bosons.

An $1/4$ BPS dyon state is generically a bound state of an electric
string and a magnetic string with charges respectively $q_e$ and
$q_m$, wrapped on the two cycles of the torus.

\begin{figure}
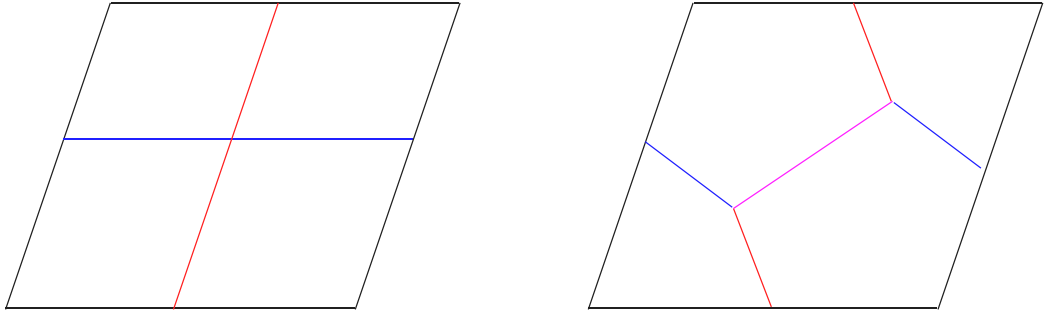

\centering \epsfig{file=four.eps, width=3in} \epsfig{file=three.eps,
width=3in}\caption{Intersection of strings of charges $q_e$(red) and
$q_m$(blue) relaxes two supersymetric junctions joined by a string
of charges $q_e + q_m$ \label{three}}
\end{figure}

The naive configuration of the two strings wrapping the two cycles
and intersecting at a four-pronged intersection is not
supersymmetric. We want to argue that it can reach a lower energy,
$1/4$ BPS ground state by relaxing the intersection and splitting it
up, to form two three-pronged string junctions joined by a stub of
charges $q_e + q_m$ or $q_e - q_m$ (See Figure \ref{three}).

In other words the dyon can be realized as a network or web of black
strings wrapping the $T^2$, made out of three strands of different
charges joined at two supersymmetric junctions.

%\begin{figure}
%\centering \caption{Non-supersymmetric intersection of strings of
%charges $q_e$(red) and $q_m$(blue) \label{four}}
%\end{figure}

\begin{figure}
\centering \epsfig{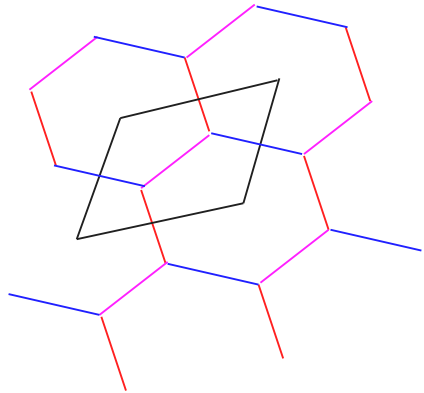} \caption{An honeycomb
network can be quotiented to give the network on the torus
\label{web}}
\end{figure}

It is well known that a junction of three BPS strings can preserve
$1/4$ of the supersymmetries. As long as the strings are in a plane,
charge is conserved at the junction and the relative angles are
fixed by mechanical equilibrium \cite{Aharony:1997bh} the BPS bound
 is saturated. This is true in particular of these black strings
from $IIB$ compactified on $K3$. Periodic, honeycomb-like networks
of strings of charges $q_e$,$q_m$,$q_e + q_m$ can be built out of
those intersections (See for example \cite{Hollowood:2003cv}).
Quotient by the periodicity of the lattice gives the $1/4$ BPS
network wrapping a $T^2$, with two three-pronged intersections, that
corresponds to a BPS dyon in four dimensions. (See Figure \ref{web})

The actual angles between the strands depend on the background
values of the scalar moduli: the $K3$ moduli fix the tensions of the
three strands and hence the relative angles at the junction, while
the shape of the torus fixes the position and orientation of the
junction. It should be possible to make use of this construction to
understand the attractor equations geometrically.

The topology of the string network wrapping the torus is the same as
a genus two Riemann surface with very thin handles. This can be made
precise: to count the (index) number of BPS microstates of such a
string network, one computes the partition function with euclidean
time compactified on a supersymmetric circle.

By taking the limit in which the supersymmetric thermal circle is
very small we can dualize IIB theory to M-theory on a torus. The
partition function is then computed in M-theory compactified on $K3
\times T^4$.

The duality relates a network of black strings wrapped on the torus
to an $M5$ brane wrapping $K3$ and a holomorphic curve in the $T^4$.
The topology of the string network with two three-pronged junctions
implies that the curve is of genus two. There is a natural
holomorphic map from a genus two Riemann surface into a $T^4$, that
is the map between the complex curve and its Jacobian. This map is
unique.

The partition function for the $M5$ brane wrapping the $K3$ and a
genus two Riemann surface is easily computed. For example by a
second duality from M-theory on $K3$ to heterotic string on $T^3$,
this is the partition function for a fundamental heterotic string
wrapping the genus two Riemann surface.

As the fermions are periodic around each circle, the partition
function is computed with Ramond boundary conditions around each
cycle of the curve, that projects the right-moving sector onto the
Ramond ground state. The electric and magnetic charges along each
strand of the network fix the value of the Narain momenta running
across the various cycles of the Riemann surface.

Level matching identifies then $N_e$ with the left oscillator level
propagating along one handle of the surface, $N_m$ along the other
handle, $N_e + N_m + N_{em}$ along the stub. The number of states
contributing to the partition function is computed from the Fourier
coefficients of the genus two partition function for $24$ chiral
bosons: $\Phi({\rho \, v \choose v \, \sigma}) = \sum
d(N_e,N_{em},N_m) e^{2 \pi i (N_e \rho + N_{em} v + N_m \sigma)}$

\begingroup\raggedright

\endgroup

\end{document}